\documentclass[11pt]{article}
\usepackage{epsfig}
\begin{document}
%%%%%%%%%%%%%%%%%%%%
\title{{\bf{\Large A New Global Embedding Approach to Study Hawking and Unruh Effects}}}
%%%%%%%%%%%%%%%%%%%%
\author{
 {\bf {\normalsize Rabin Banerjee}$
$\thanks{E-mail: rabin@bose.res.in}},\, 
 {\bf {\normalsize Bibhas Ranjan Majhi}$
$\thanks{E-mail: bibhas@bose.res.in}}\\
 {\normalsize S.~N.~Bose National Centre for Basic Sciences,}
\\{\normalsize JD Block, Sector III, Salt Lake, Kolkata-700098, India}
\\[0.3cm]
}
%\date{}

\maketitle

\begin{abstract}
      A new type of global embedding of curved space-times in higher dimensional flat ones is introduced to present a unified description of Hawking and Unruh effects. Our analysis simplifies as well as generalises the conventional embedding approach.     
\end{abstract}

\section{Introduction}
         After Hawking's famous work \cite{Hawking:1974rv} - the black holes radiate - known as {\it{Hawking effect}}, it is now well understood that it is related to the event horizon of a black hole. A closely related effect is the {\it{Unruh effect}} \cite{Unruh:1976db}, where a similar type of horizon is experienced by a uniformly accelerated observer on the Minkowski space-time. A unified description of them was first put forwarded by Deser and Levin \cite{Deser:1997ri,Deser:1998xb} which was a sequel to an earlier attempt \cite{Narnhofer:1996zk}. This is called the global embedding Minkowskian space (GEMS) approach. In this approach, the relevant detector in curved space-time (namely Hawking detector) and its event horizon map to the Rindler detector in the corresponding flat higher dimensional embedding space \cite{Goenner,Rosen} and its event horizon. 
%They showed that the Hawking effect on a curved manifold can be looked at as the Unruh effect in a higher dimensional flat space.
%It was well known that a curved manifold can be globally embedded to a higher dimensional space, popularly known as GEMS \cite{}. Using this idea Deser-Levin \cite{} showed that it is possible to establish a mapping between the detector in the curved space-time (Hawking detector) and a detector (Rindler detector) in the corresponding flat higher dimensional embedding space. 
Then identifying the acceleration of the Unruh detector, the Unruh temperature was calculated. Finally, use of the Tolman relation \cite{Tolman} yields the Hawking temperature. Subsequently, this unified approach to determine the Hawking temperature using the Unruh effect was applied for several black hole space-times \cite{Kim:2000ct,Tian:2005yj,Brynjolfsson:2008uc,Hong:2003xz}. However the results were confined to four dimensions and the calculations were done case by case, taking specific black hole metrics. It was not clear whether the technique was applicable to complicated examples like the Kerr-Newman metric which lacks spherical symmetry.
% Finally, an explicit derivation of the spectrum of emitted particles from the event horizon was also not done within this approach. We feel this to be a serious omission in the GEMS approach.      

      The motivation of this paper is to give a modified presentation of the GEMS approach that naturally admits generalization. Higher dimensional black holes with different metrics, including Kerr-Newman, are considered. 
Using this new embedding, the local Hawking temperature (Unruh temperature) will be derived. Then the Tolman formula leads to the Hawking temperature. 
%This will thereby give a complete description of unifying the Hawking and Unruh effects by the GEMS approach.

      We shall first introduce a new global embedding which embeds only the ($t-r$)-sector of the curved metric into a flat space. It will be shown that this embedding is enough to derive the Hawking result using the Deser-Levin approach \cite{Deser:1997ri,Deser:1998xb}, instead of the full embedding of the curved space-time. Hence we might as well call this the reduced global embedding. This is actually motivated from the fact that an $N$-dimensional black hole metric effectively reduces to a $2$ -dimensional metric (only the ($t-r$)-sector) near the event horizon by the dimensional reduction technique \cite{Robinson:2005pd,Carlip:1998wz,Iso:2006ut,Umetsu:2009ra} (for examples see Appendix 1). Furthermore, this $2$-dimensional metric is enough to find the Hawking quantities if the back scattering effect is ignored. Several spherically symmetric static metrics will be exemplified. Also, to show the utility of this reduced global embedding, we shall discuss the most general solution of the Einstein gravity - Kerr-Newman space-time, whose full global embedding is difficult to find. Since the reduced embedding involves just the two dimensional ($t-r$)-sector, black holes in arbitrary dimensions can be treated. 
In this sense our approach is valid for any higher dimensional black hole. 

%     The Unruh emission spectrum will be derived by the reformulated tunneling approach \cite{Banerjee:2008sn,Banerjee:2009wb,Banerjee:2009pf,Roy:2009vy}, which is different from the usual tunneling approach \cite{Srinivasan:1998ty,Parikh:1999mf,Banerjee:2008cf}. For that we shall first find the relevant Rindler metric using the reduced global embedding for different black holes. Then considering the propagation of scalar particles under this background the required relations between the inside modes and the outside modes will be established. Using a density matrix technique \cite{Banerjee:2009pf}, the Unruh spectrum will be calculated, which has the desired perfect black body structure. The Unruh temperature will be immediately identified from the Boltzmann factor. The well known Hawking temperature/spectrum is then easily found out by using the Tolman relation \cite{Tolman}. 

     The organization of the paper is as follows. In section 2 we shall find the reduced global embedding of several black hole space-times which are spherically symmetric. In the next section the power of this approach will be exploited to find the Unruh/Hawking temperature for the Kerr-Newman black hole. Finally, we shall give our concluding remarks. 

\section{Reduced global embedding}
    A unified picture of Hawking effect \cite{Hawking:1974rv} and Unruh effect \cite{Unruh:1976db} was established by the global embedding of a curved space-time into a higher dimensional flat space \cite{Deser:1998xb}. %In this approach the relevant detector in curved space-time (namely Hawking detector) and its event horizon map to the Rindler detector in the corresponding flat higher dimensional embedding space and its event horizon. 
Subsequently, this unified approach to determine the Hawking temperature using the Unruh effect was applied for several black hole space-times \cite{Kim:2000ct,Tian:2005yj}, but usually these are spherically symmetric. For instance, no discussion on the Kerr-Newman black hole has been given, because it is difficult to find the full global embedding.

     Since the Hawking effect is governed solely by properties of the event horizon, it is enough to consider the near horizon theory. As already stated, this is a two dimensional theory obtained by dimensional reduction of the full theory. Its metric is just the ($t-r$)-sector of the original metric.

   In the following sub-sections we shall find the global embedding of the near horizon effective $2$-dimensional theory. Then the usual local Hawking temperature will be calculated. Technicalities are considerably simplified and our method is general enough to include different black hole metrics. 

\subsection{Schwarzschild metric}
   Near the event horizon the physics is given by just the two dimensional ($t-r$) -sector of the full Schwarzschild metric \cite{Robinson:2005pd}:
\begin{eqnarray}
ds^2 = g_{tt}dt^2 + g_{rr}dr^2 = \Big(1-\frac{2m}{r}\Big)dt^2 -\frac{dr^2}{1-\frac{2m}{r}}.
\label{1.04}
\end{eqnarray} 
It is interesting to see that this can be globally embedded in a flat $D=3$ space as,
\begin{eqnarray}
ds^2 = (dz^0)^2 - (dz^1)^2 - (dz^2)^2
\label{1.32}
\end{eqnarray}
by the following relations among the flat and curved coordinates:
\begin{eqnarray}
&&z^0_{out} = \kappa^{-1} \Big(1-\frac{2m}{r}\Big)^{1/2} \textrm{sinh}(\kappa t),\,\,\,\
z^1_{out} = \kappa^{-1} \Big(1-\frac{2m}{r}\Big)^{1/2} \textrm{cosh}(\kappa t),
\nonumber
\\
&&z^0_{in} = \kappa^{-1} \Big(\frac{2m}{r} - 1\Big)^{1/2} \textrm{cosh}(\kappa t),\,\,\,\
z^1_{in} = \kappa^{-1} \Big(\frac{2m}{r} - 1\Big)^{1/2} \textrm{sinh}(\kappa t),
\nonumber
\\
&&z^2 = \int dr \Big(1+\frac{r_Hr^2 + r_H^2r + r_H^3}{r^3}\Big)^{1/2},
\label{1.33}
\end{eqnarray}
where the surface gravity $\kappa=\frac{1}{4m}$ and the event horizon is located at $r_H=2m$. The suffix ``$in$'' (``$out$'') refer to the inside (outside) of the event horizon while variables without any suffix imply that these are valid on both sides of the horizon. We shall follow these notations throughout the paper. Now if a detector moves according to constant $r$ (Hawking detector) outside the horizon in the curved space, then the corresponding Unruh detector moves on the constant $z^2$ plane and it will follow the hyperbolic trajectory
\begin{eqnarray}
\Big(z^1_{out}\Big)^2 - \Big(z^0_{out}\Big)^2 = 16 m^2 \Big(1 - \frac{2m}{r}\Big) = \frac{1}{{\tilde{a}}^2}.
\label{hyper}
\end{eqnarray}
This shows that the Unruh detector is moving in the ($z^0_{out}, z^1_{out}$) flat plane with a uniform acceleration ${\tilde{a}}= \frac{1}{4m}\Big(1 - \frac{2m}{r}\Big)^{-1/2}$. Then, according to Unruh \cite{Unruh:1976db}, the accelerated detector will see a thermal spectrum in the Minkowski vacuum with the local Hawking temperature given by,
\begin{eqnarray}
T = \frac{\hbar {\tilde{a}}}{2\pi} = \frac{\hbar}{8\pi m} \Big(1 - \frac{2m}{r}\Big)^{-1/2}.
\label{localtemp}
\end{eqnarray}
So we see that with the help of the reduced global embedding the local Hawking temperature near the horizon can easily be obtained.

    Now the temperature measured by any observer away from the horizon can be obtained by using the Tolman formula \cite{Tolman} which ensures constancy between the product of temperatures and corresponding Tolman factors measured at two different points in space-time. This formula is given by \cite{Tolman}:
\begin{eqnarray}
\sqrt{g_{tt}}~ T = \sqrt{g_{0_{tt}}}~ T_0
\label{new1}
\end{eqnarray}
where, in this case, the quantities on the left hand side are measured near the horizon whereas those on the right hand side are measured away from the horizon (say at $r_0$). Since away from the horizon the space-time is given by the full metric, $g_{0_{tt}}$ must correspond to the $dt^2$ coefficient of the full (four dimensional) metric.

   For the case of Schwarzschild metric $g_{tt} = 1-2m/r$, $g_{0_{tt}} = 1-2m/r_0$. Now the Hawking effect is observed at infinity ($r_0 = \infty$), where $g_{0_{tt}} = 1$. Hence, use of the Tolman formula (\ref{new1}) immediately yields the Hawking temperature:
\begin{eqnarray}
T_0 = {\sqrt{g_{tt}}}~ T =  \frac{\hbar}{8\pi m }.
\label{hawkingtemp}
\end{eqnarray}
Thus, use of the reduced embedding instead of the embedding of the full metric is sufficient to get the answer. 
%Now following the usual Deser-Levin approach \cite{Deser:1998xb} Hawking and Unruh effects get unified.

\subsection {Reissner-Nordstr$\ddot{\textrm{o}}$m metric}
   In this case, the effective metric near the event horizon is given by \cite{Robinson:2005pd},
\begin{eqnarray}
ds^2 = \Big(1 - \frac{2m}{r} + \frac{e^2}{r^2}\Big)dt^2 - \frac{dr^2}{1 - \frac{2m}{r} + \frac{e^2}{r^2}}.
\label{1.34}
\end{eqnarray}
This metric can be globally embedded into the $D=4$ dimensional flat metric as,
\begin{eqnarray}
ds^2 = (dz^0)^2 - (dz^1)^2 - (dz^2)^2 + (dz^3)^2
\label{1.35}
\end{eqnarray}
where the coordinate transformations are:
\begin{eqnarray}
&&z^0_{out} = \kappa^{-1} \Big(1-\frac{2m}{r} + \frac{e^2}{r^2}\Big)^{1/2} \textrm{sinh}(\kappa t),\,\,\,\
z^1_{out} = \kappa^{-1} \Big(1-\frac{2m}{r} + \frac{e^2}{r^2}\Big)^{1/2} \textrm{cosh}(\kappa t),
\nonumber
\\
&&z^0_{in} = \kappa^{-1} \Big(\frac{2m}{r} - \frac{e^2}{r^2} - 1\Big)^{1/2} \textrm{cosh}(\kappa t),\,\,\,\
z^1_{in} = \kappa^{-1} \Big(\frac{2m}{r} - \frac{e^2}{r^2} - 1\Big)^{1/2} \textrm{sinh}(\kappa t),
\nonumber
\\
&&z^2 = \int dr \Big[1+\frac{r^2(r_+ + r_-) + r_+^2(r + r_+)}{r^2(r-r_-)}\Big]^{1/2},
\nonumber
\\
&&z^3 = \int dr \Big[\frac{4r_+^5r_-}{r^4(r_+ - r_-)^2}\Big]^{1/2}.
\label{1.36}
\end{eqnarray}
Here in this case the surface gravity $\kappa = \frac{r_+ - r_-}{2r_+^2}$ and $r_{\pm}=m\pm\sqrt{m^2-e^2}$. The black hole event horizon is given by $r_H=r_+$. Note that for $e=0$, the above transformations reduce to the Schwarzschild case (\ref{1.33}). 
%Now, again following Deser-Levin approach \cite{Deser:1998xb}, 
The Hawking detector moving in the curved space outside the horizon, following a constant $r$ trajectory, maps to the Unruh detector on the constant ($z^2,z^3$) surface. The trajectory of the Unruh detector is given by
\begin{eqnarray}
\Big(z^1_{out}\Big)^2 - \Big(z^0_{out}\Big)^2 = \Big(\frac{r_+ - r_-}{2r_+^2}\Big)^{-2} \Big(1-\frac{2m}{r} + \frac{e^2}{r^2}\Big)=\frac{1}{{\tilde{a}}^2}.
\label{RN1}
\end{eqnarray}
This, according to Unruh \cite{Unruh:1976db}, immediately leads to the local Hawking temperature $T=\frac{\hbar {\tilde{a}}}{2\pi}=\frac{\hbar(r_+ - r_-)}{4\pi r_+^2\sqrt{1-2m/r+e^2/r^2}}$ which was also obtained from the full global embedding \cite{Deser:1998xb}. Again, since in this case $g_{0_{tt}} = 1-2m/r_0 + e^2/r_0^2$ which reduces to unity at $r_0=\infty$ and $g_{tt} = 1-2m/r+e^2/r^2$, use of Tolman formula (\ref{new1}) leads to the standard Hawking temperature $T_0=\sqrt{g_{tt}}~ T=\frac{\hbar(r_+ - r_-)}{4\pi r_+^2}$.

\subsection{Schwarzschild-AdS metric}
   Near the event horizon the relevant effective metric is \cite{Robinson:2005pd},
\begin{eqnarray}
ds^2 = \Big(1-\frac{2m}{r}+\frac{r^2}{R^2}\Big)dt^2 - \frac{dr^2}{\Big(1-\frac{2m}{r}+\frac{r^2}{R^2}\Big)},
\label{1.37}
\end{eqnarray}
where $R$ is related to the cosmological constant $\Lambda= -1/R^2$.
This metric can be globally embedded in the flat space (\ref{1.35}) with the following coordinate transformations:
\begin{eqnarray}
&&z^0_{out} = \kappa^{-1} \Big(1-\frac{2m}{r} + \frac{r^2}{R^2}\Big)^{1/2} \textrm{sinh}(\kappa t),\,\,\
z^1_{out} = \kappa^{-1} \Big(1-\frac{2m}{r} + \frac{r^2}{R^2}\Big)^{1/2} \textrm{cosh}(\kappa t),
\nonumber
\\
&&z^0_{in} = \kappa^{-1} \Big(\frac{2m}{r} - \frac{r^2}{R^2} - 1\Big)^{1/2} \textrm{cosh}(\kappa t),\,\,\,\
z^1_{in} = \kappa^{-1} \Big(\frac{2m}{r} - \frac{r^2}{R^2} - 1\Big)^{1/2} \textrm{sinh}(\kappa t),
\nonumber
\\
&&z^2 = \int dr \Big[1+\Big(\frac{R^3 + R r_H^2}{R^2 + 3r_H^2}\Big)^2 \frac{r^2r_H+rr_H^2 + r_H^3}{r^3(r^2+rr_H+r_H^2+R^2)}\Big]^{1/2},
\nonumber
\\
&&z^3 = \int dr \Big[\frac{(R^4 + 10R^2r_H^2+9r_H^4)(r^2+rr_H+r_H^2)}{(r^2+rr_H+r_H^2+R^2)(R^2+3r_H^2)^2}\Big]^{1/2}
\label{1.38}
\end{eqnarray}
where the surface gravity $\kappa=\frac{R^2+3r_H^2}{2r_HR^2}$ and the event horizon $r_H$ is given by the root of the equation $1-\frac{2m}{r_H} + \frac{r^2_H}{R^2}=0$.
Note that in the $R\rightarrow\infty$ limit these transformations reduce to those for the Schwarzschild case (\ref{1.33}). We observe that the Unruh detector on the ($z^2,z^3$) surface (i.e. the Hawking detector moving outside the event horizon on a constant $r$ surface) follows the hyperbolic trajectory:
\begin{eqnarray}
\Big(z^1_{out}\Big)^2 - \Big(z^0_{out}\Big)^2 = \Big(\frac{R^2+3r_H^2}{2r_HR^2}\Big)^{-2}\Big(1-\frac{2m}{r} + \frac{r^2}{R^2}\Big)=\frac{1}{{\tilde{a}}^2}
\label{ADS1}
\end{eqnarray} 
leading to the local Hawking temperature $T=\frac{\hbar {\tilde{a}}}{2\pi}=\frac{\hbar\kappa}{2\pi\Big(1-\frac{2m}{r}+\frac{r^2}{R^2}\Big)^{1/2}}$. This result was obtained earlier \cite{Deser:1998xb}, but with more technical complexities,  from the embedding of the full metric.

      It may be pointed out that for the present case, the observer must be at a finite distance away from the event horizon, since the space-time is asymptotically AdS. Therefore, if the observer is far away from the horizon ($r_0>>r$) where $g_{0_{tt}}=1-2m/r_0+r_0^2/R^2$, then use of (\ref{new1}) immediately leads to the temperature measured at $r_0$:
\begin{eqnarray}
T_0 = \frac{\hbar\kappa}{2\pi\sqrt{1-2m/r_0 + r_0^2/R^2}}.
\label{new2}
\end{eqnarray}     
Now, this shows that $T_0\rightarrow 0$ as $r_0\rightarrow \infty$; i.e. no Hawking particles are present far from horizon. 
%This is because they do not have enough energy to escape to infinity where the effective potential is infinity.

\section{Kerr-Newman metric}
           So far we have discussed a unified picture of Unruh and Hawking effects using our reduced global embedding approach for spherically symmetric metrics, reproducing standard results. However, our approach was technically simpler since it involved the embedding of just the two dimensional near horizon metric. Now we shall explore the real power of this new embedding.

           The utility of the reduced embedding approach comes to the fore for the Kerr-Newman black hole which is not spherically symmetric. The embedding for the full metric, as far as we are aware, is not done in the literature.

   The effective $2$-dimensional metric near the event horizon is given by \cite{Iso:2006ut,Umetsu:2009ra},
\begin{eqnarray}
ds^2 = \frac{\Delta}{r^2+a^2}dt^2 - \frac{r^2+a^2}{\Delta}dr^2,
\label{Kerr1}
\end{eqnarray} 
where 
\begin{eqnarray}
&&\Delta = r^2-2mr+a^2+e^2 = (r-r_+)(r-r_-);\,\,\,\ a=\frac{J}{m};
\nonumber
\\
&&r_\pm = m\pm\sqrt{m^2-a^2-e^2}.
\label{Kerr2}
\end{eqnarray}
The event horizon is located at $r=r_+$. This metric can be embedded in the following $D=5$-dimensional flat space:
\begin{eqnarray}
ds^2=\Big(dz^0\Big)^2 -\Big(dz^1\Big)^2-\Big(dz^2\Big)^2 + \Big(dz^3\Big)^2 + \Big(dz^4\Big)^2,
\label{Kerr3}
\end{eqnarray}
where the coordinate transformations are
\begin{eqnarray}
&&z^0_{out} = \kappa^{-1} \Big(1-\frac{2mr}{r^2+a^2} + \frac{e^2}{r^2+a^2}\Big)^{1/2} \textrm{sinh}(\kappa t),\,\,\,\
z^1_{out} = \kappa^{-1} \Big(1-\frac{2mr}{r^2+a^2} + \frac{e^2}{r^2+a^2}\Big)^{1/2} \textrm{cosh}(\kappa t),
\nonumber
\\
&&z^0_{in} = \kappa^{-1} \Big(\frac{2mr}{r^2+a^2} - \frac{e^2}{r^2+a^2} - 1\Big)^{1/2} \textrm{cosh}(\kappa t),\,\,\,\
z^1_{in} = \kappa^{-1} \Big(\frac{2mr}{r^2+a^2} - \frac{e^2}{r^2+a^2} - 1\Big)^{1/2} \textrm{sinh}(\kappa t),
\nonumber
\\
&&z^2 = \int dr \Big[1+\frac{(r^2+a^2)(r_+ + r_-) + r_+^2(r + r_+)}{(r^2+a^2)(r-r_-)}\Big]^{1/2},
\nonumber
\\
&&z^3 = \int dr \Big[\frac{4r_+^5r_-}{(r^2+a^2)^2(r_+ - r_-)^2}\Big]^{1/2},
\nonumber
\\
&&z^4 = \int dr a\Big[\frac{r_+ + r_-}{(a^2+r_-^2)(r_- - r)} + \frac{4(a^2 + r_+^2)(a^2-r_+r_- + (r_+ +r_-)r)}{(r_+ - r_-)^2 (a^2 + r^2)^3} 
\nonumber
\\
&&+ \frac{4r_+r_-(a^2+2r_+^2)}{(r_+ - r_-)^2(a^2+r^2)^2} + \frac{rr_- - a^2 + r_+(r+r_-)}{(a^2+r_-^2)(a^2+r^2)}\Big]^{1/2}.
\label{Kerr4}
\end{eqnarray}
Here the surface gravity $\kappa = \frac{r_+-r_-}{2(r_+^2 + a^2)}$.
For $e=0, a=0$, as expected, the above transformations reduce to the Schwarzschild case (\ref{1.33}) while only for $a=0$ these reduce to the Reissner-Nordstr$\ddot{\textrm{o}}$m case (\ref{1.36}).

    As before, the trajectory adopted by the Unruh detector on the constant ($z^2,z^3,z^4$) surface corresponding to the Hawking detector on the constant $r$ surface is given by the hyperbolic form,
\begin{eqnarray}
\Big(z^1_{out}\Big)^2 - \Big(z^0_{out}\Big)^2 = \kappa^{-2}\Big(1-\frac{2mr}{r^2+a^2} + \frac{e^2}{r^2+a^2}\Big)=\frac{1}{{\tilde{a}}^2}.
\label{Kerr5}
\end{eqnarray}
Hence the local Hawking temperature is
\begin{eqnarray}
T=\frac{\hbar {\tilde{a}}}{2\pi}=\frac{\hbar\kappa}{2\pi\sqrt{\Big(1-\frac{2mr}{r^2+a^2} + \frac{e^2}{r^2+a^2}\Big)}}.
\label{Kerr6}
\end{eqnarray} 
Finally, since $g_{tt} = 1-\frac{2mr}{r^2+a^2} + \frac{e^2}{r^2+a^2}$ (corresponding to the near horizon reduced two dimensional metric) and $g_{0_{tt}}=\frac{r_0^2-2mr_0+a^2+e^2-a^2 {\textrm{sin}}^2\theta}{r_0^2+a^2{\textrm{cos}}^2\theta}$ (corresponding to the full four dimensional metric), use of the Tolman relation (\ref{new1}) leads to the Hawking temperature
\begin{eqnarray}
T_0=\frac{\sqrt{g_{tt}}}{\sqrt{(g_0{_{tt}})_{r_0\rightarrow\infty}}}~T = \frac{\hbar\kappa}{2\pi} = \frac{\hbar(r_+-r_-)}{4\pi(r_+^2 + a^2)},
\label{Kerr7}
\end{eqnarray}
which is the well known result \cite{Iso:2006ut}.

\section{Conclusion}
      We provide a new approach to the study of Hawking/Unruh effects including their unification, initiated in \cite{Deser:1997ri,Deser:1998xb,Narnhofer:1996zk}, popularly known as global embedding Minkowskian space-time (GEMS). Contrary to the usual formulation \cite{Deser:1997ri,Deser:1998xb,Narnhofer:1996zk,Kim:2000ct,Tian:2005yj,Brynjolfsson:2008uc}, the full embedding was avoided. Rather, we required the embedding of just the two dimensional ($t-r$)-sector of the theory. This was a consequence of the fact that the effective near horizon theory is basically two dimensional. Only near horizon theory is significant since Hawking/Unruh effects are governed solely by properties of the event horizon.

    This two dimensional embedding ensued remarkable technical simplifications whereby the treatment of more general black holes (e.g. those lacking spherical symmetry like the Kerr-Newman) was feasible. Also, black holes in any dimensions were automatically considered since the embedding just required the ($t-r$)-sector.

\end{document}